\documentclass[9pt,twocolumn,twoside]{osajnl}

\journal{ol}
\setboolean{shortarticle}{true}

\title{Acoustic Phonon Sideband Dynamics During Polaron Formation in a Single Quantum Dot}
\author[1,6]{Daniel Wigger}
\author[2]{Vage Karakhanyan}
\author[3]{Christian Schneider}
\author[3]{Martin Kamp}
\author[3,4]{Sven~H\"ofling}
\author[5]{Pawe\l{}~Machnikowski}
\author[1]{Tilmann~Kuhn}
\author[2,6]{Jacek Kasprzak}
\affil[1]{Institute of Solid State Theory, University of M\"unster, Wilhelm-Klemm-Str. 10, 48149 M\"unster, Germany}
\affil[2]{Universit'e Grenoble Alpes, CNRS, Grenoble INP, Institut N\'{e}el, 38000 Grenoble, France}
\affil[3]{Technische Physik, Universit\"at W\"urzburg, 97074 W\"urzburg, Germany}
\affil[4]{SUPA, School of Physics and Astronomy, University of St. Andrews, St. Andrews KY16 9SS, UK}
\affil[5]{Department of Theoretical Physics, Wroc\l{}aw University of Science and Technology, 50-370 Wroc\l{}aw, Poland}
\affil[6]{e-mail: d.wigger@wwu.de, jacek.kasprzak@neel.cnrs.fr}
\begin{abstract}
When an electron-hole pair is optically excited in a semiconductor
quantum dot the host crystal lattice needs
to adapt to the presence of the generated charge distribution. Therefore the coupled
exciton-phonon system has to establish a new equilibrium, which is
reached in the form of a quasiparticle called polaron. Especially,
when the exciton is abruptly generated on a timescale faster than
the typical lattice dynamics, the lattice displacement cannot follow
adiabatically. Consequently, a rich dynamics on the picosecond
timescale of the coupled system is expected. In this study we
combine simulations and measurements of the ultrafast, coherent,
nonlinear optical response, obtained by four-wave mixing spectroscopy, to resolve the
formation of this polaron. By detecting and investigating the phonon
sidebands in the four-wave mixing spectra for varying pulse delays
and different temperatures we have access to the influence of
phonon emission and absorption processes which finally result in the
emission of an acoustic wave packet out from the quantum dot.
\end{abstract}

\setboolean{displaycopyright}{true}

\begin{document}

\maketitle
\begin{figure}[t]
\centering \fbox{\includegraphics[width=0.82\linewidth]{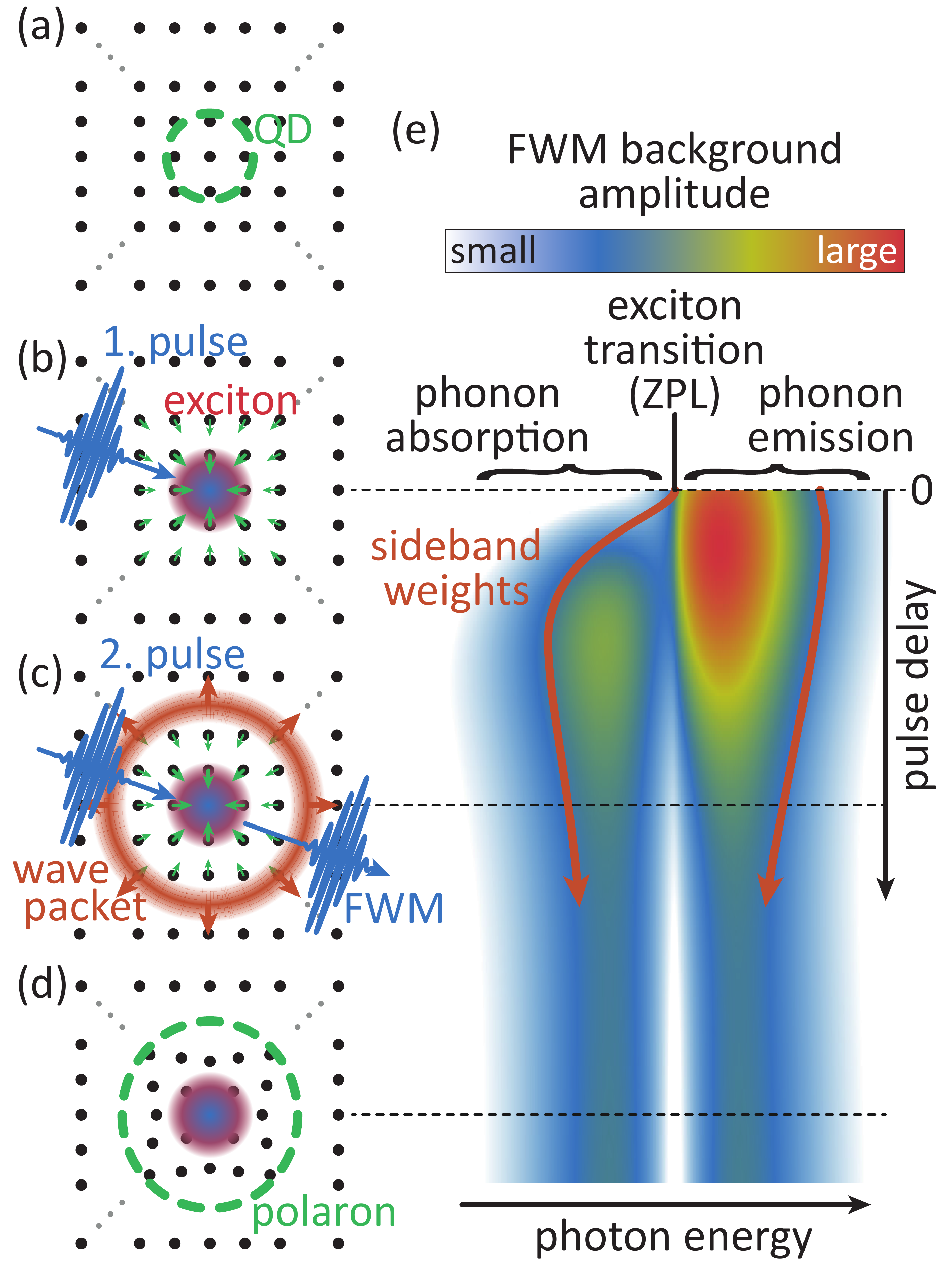}}
\caption{Schematic picture of the polaron generation and its
measurement. (a) Empty QD system in equilibrium. (b) The laser pulse
creates an exciton in the QD. The charge distributions attracts the
surrounding lattice atoms. (c) The fast contraction of the lattice
leads to the emission of a phonon wave packet. The second laser
pulse is used to probe the exciton dynamics. (d) New equilibrium
state in the presence of an exciton is called polaron. (e) Phonon
background spectrum of the FWM signal. The delay increases from top
to bottom. The dashed lines loosely correspond to the schematics in
(b), (c), and(d). The orange arrows follow the intensity
distribution of the two sidebands.} \label{fig:1}
\end{figure}
\begin{figure}[t]
\centering
\fbox{\includegraphics[width=0.85\linewidth]{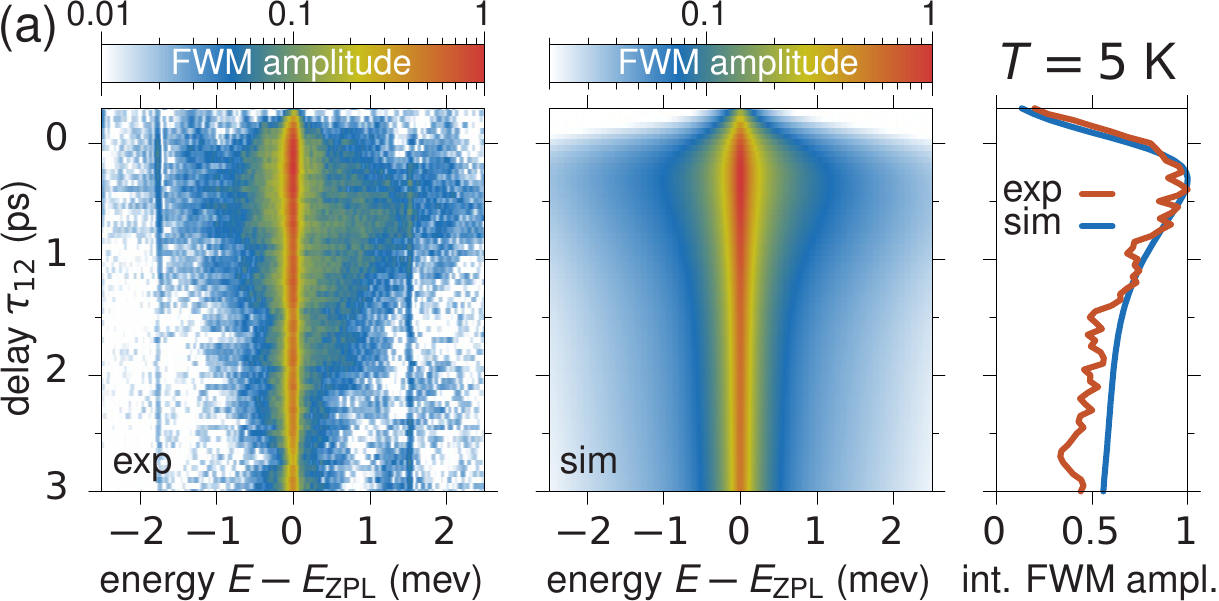}}
\fbox{\includegraphics[width=0.85\linewidth]{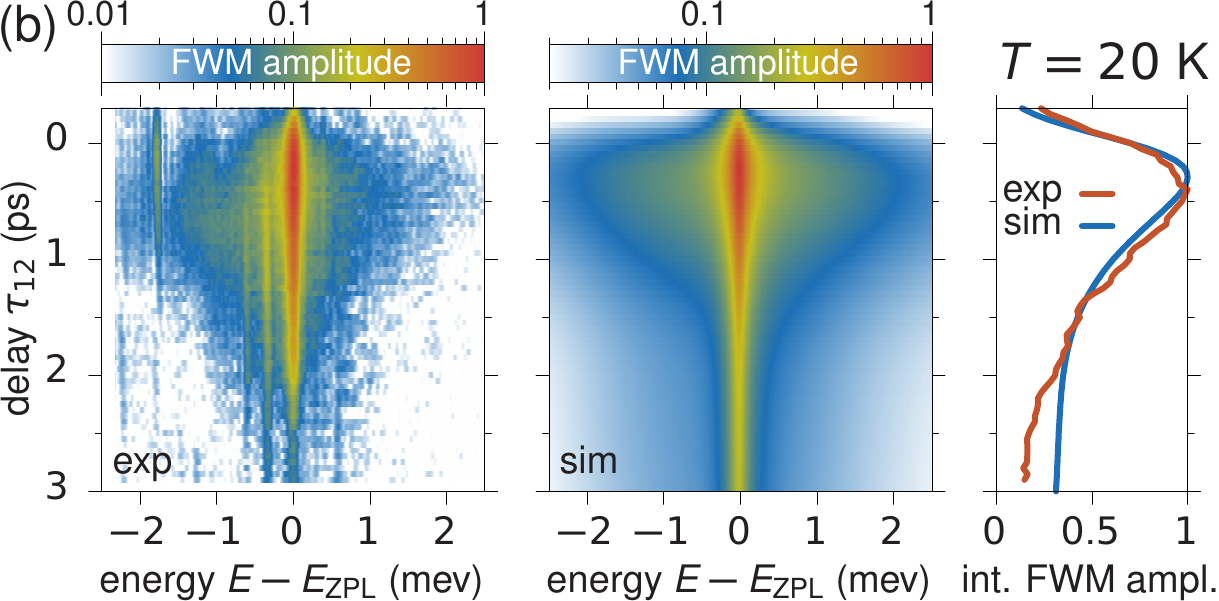}}
\fbox{\includegraphics[width=0.85\linewidth]{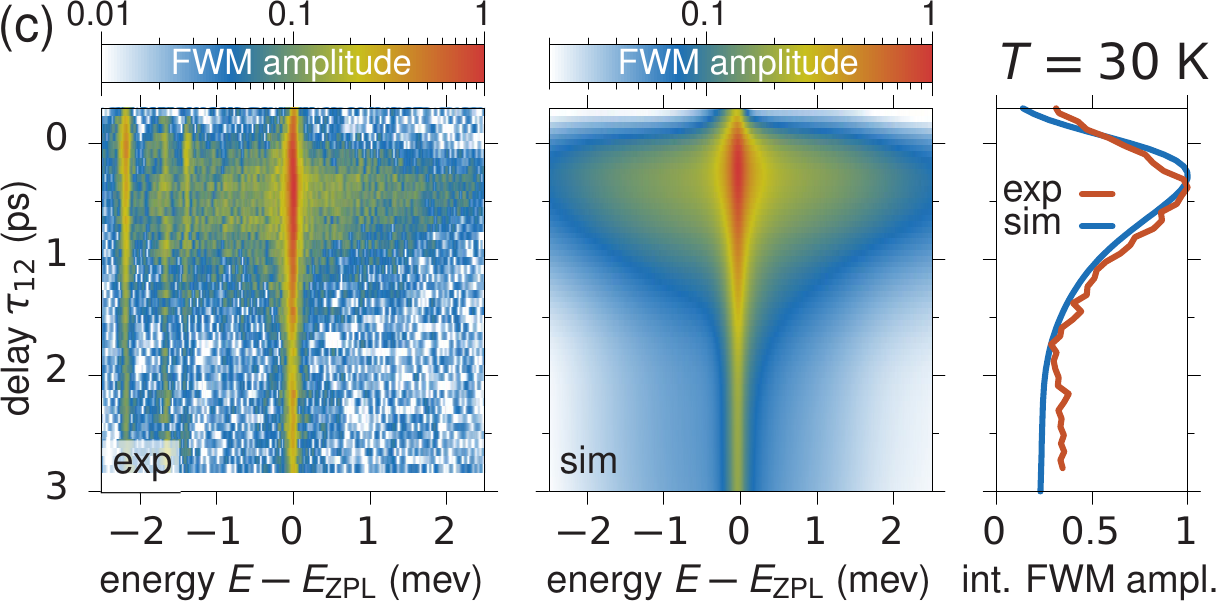}}
\caption{Measured and simulated FWM spectra for varying pulse delay $\tau_{12}$ left and center, respectively. The corresponding spectrally integrated and normalized FWM amplitudes are shown on the right, experiment in orange and theory in blue. The temperatures are $T=5$~K in (a), $T=20$~K in (b), and $T=30$~K in (c).}
\label{fig:2}
\end{figure}
\begin{figure}[t]
\centering
\fbox{\includegraphics[width=0.85\linewidth]{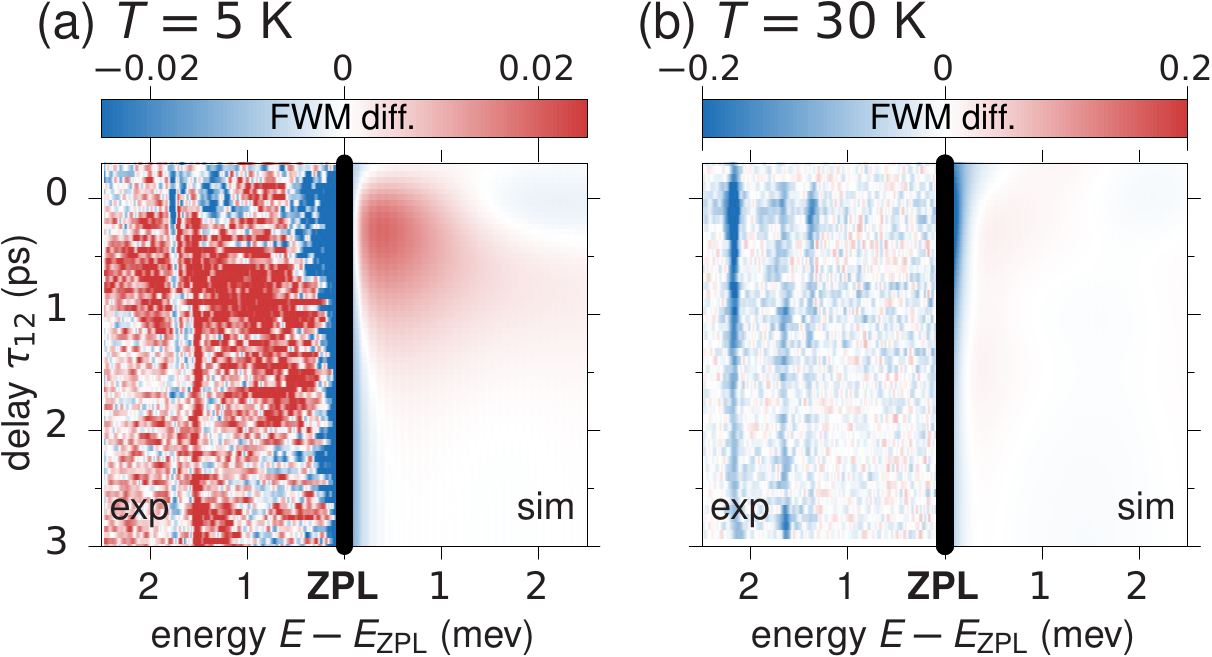}}
\caption{Difference between phonon emission and phonon absorption side of the FWM spectra for (a) $T=5$~K and (b) $T=30$~K. The difference is calculated by subtracting the FWM amplitude for $E-E_{\rm ZPL}<0$ from the one at $E-E_{\rm ZPL}>0$. The experiment is shown on the left and the simulation on the right, respectively. To isolate the influence of the phonon sidebands the black line blocks the ZPL influence.}
\label{fig:3}
\end{figure}

The coupling between quantum dot (QD) excitons and acoustic phonons is unavoidable in self-assembled systems. On the one hand the dephasing effect of the lattice vibrations is in general seen as detrimental. On the other hand many new approaches focus on the active use of phonons in nanophotonics. Phonon assisted transitions can be used for deterministic state preparation~\cite{reiter2014role} and even information transfer~\cite{huneke2008impa,lemonde2018phon}. Exciton-phonon coupling can also serve as an interface to induce and optically control the quantum state of localized phonon modes~\cite{reiter2011gene} or even the macroscopic motion of mechanical oscillators~\cite{wilson2004lase,auffeves2014opti,hahn2019infl}.\\
One important effect associated with this fundamental interaction is that by the generation of an exciton the combined charge-phonon system is brought out of equilibrium. By the generation of a localized lattice distortion it reaches a new state of rest. This new eigenstate of the entire system is called a {\it polaron}. Due to its fundamental character its investigation has a long history~\cite{frohlich1954elec,feynman1955slow}. When the exciton creation happens significantly faster than the polaron formation the process is accompanied by the emission of a phonon wave packet~\cite{wigger2014ener}. Spectrally this phonon assisted type of event appears as characteristic sidebands also in photoluminescence~\cite{besombes2001acou,peter2004phon,palinginis2004exci} and absorption~\cite{lindwall2007zero} signals in QDs and other single photon emitters~\cite{kumar2016reso,wigger2019phon}. However, the dynamics of this transition into a new equilibrium has only rarely been studied on single QDs~\cite{jakubczyk2016impa,vanacore2017ultr}. Previously, the exciton dephasing on a picosecond time scale due to acoustic phonons was shown on ensembles of QDs~\cite{vagov2004nonm,borri2005exci} which, due to the inhomogeneous broadening, does not allow for a spectrally resolved analysis.\\
We here focus on the back-action of the polaron formation onto the
excitons' coherent spectra, which are provided by four-wave mixing
(FWM) spectroscopy realized with heterodyne spectral
interferometry~\cite{langbein2006hete}. Applying a two pulse
excitation with a variable delay $\tau_{12}$ within the pair of
pulses, directly measures the evolution of the microscopic exciton
polarization, i.e., its coherence. By this approach we have access
to the spectral and time domain at the same time.\\
A schematic picture of our study is shown in
Figs.~\ref{fig:1}(a)-(d). The investigated system is a single
self-assembled (In)GaAs semiconductor QD embedded in a planar
semiconductor microcavity~\cite{maier2014brig,fras2016multi}. Its low
quality factor of around 160 allows for a spectral matching with
150~fs chirp-corrected laser pulses. Simultaneously, owing to the
enhancement of the intra-cavity field, the external resonant
excitation can be decreased significantly, resulting in record
signal-to-noise ratios for single QD FWM signals. Such an efficient
FWM retrieval was a key aspect in current experiments. It enabled
detection of phonon sidebands, with field amplitudes more than an order
of magnitude below that of the zero phonon line transition.\\
Prior to optical excitation the QD is in its ground state. 
The QD is embedded into the host crystal and is distinguished from it 
by a slightly different lattice constant. However, these details of the 
atomic structure are not important for this study, therefore we simply
mark a part of the schematic crystal
lattice in Fig.~\ref{fig:1}(a) as region of the QD by the green
dashed circle.
\\
The first exciting laser pulse resonantly creates an
excitonic polarization in the QD as depicted in Fig.~\ref{fig:1}(b). This
excitation creates a charge distribution which is
felt by the surrounding atoms making them move towards the QD
center. When the exciton generation happens on a timescale faster
than the typical timescale of the phonons, i.e. approximately below 1~picosecond, it is accompanied by the
emission of a phonon wave packet as sketched in
Fig.~\ref{fig:1}(c)~\cite{wigger2014ener}. In the final eigenstate, the coupled
exciton-phonon system forms a polaron consisting of the exciton
in the QD and a distorted lattice in the region of the dot. This is
illustrated in Fig.~\ref{fig:1}(d). To probe the back-action of
the polaron formation onto the exciton, after a delay time $\tau_{12}$ the system is excited by a
second laser pulse. The arrival time of this second pulse is varied
to measure the exciton coherence dynamics. As depicted in (c), the
second pulse creates the FWM signal carrying information on the
microscopic polarization of the exciton.\\
Assuming a pure-dephasing type coupling to acoustic phonons, the FWM signal can be obtained analytically in the limit of
ultrafast laser pulses treated as delta functions. For this purpose, the optical
excitation with two pulses carrying phases $\varphi_1$ and
$\varphi_2$ of the excitonic two level system is calculated. As
shown in Ref.~\cite{vagov2003impa}, the microscopic polarization of
the exciton filtered with respect to the FWM phase $\varphi_{\rm
FWM}=2\varphi_2-\varphi_1$ reads
\begin{eqnarray}
&&p_{\rm FWM}(t,\tau_{12})  \sim
 \exp\Bigg\{ \sum_{\bf q} \left| \frac{g_{\bf q}}{\omega_q}\right|^2
\Bigg[2\cos(\omega_q t) - 3 \\
&&\qquad + e^{i\omega_q\tau_{12}} \left(2-e^{i\omega_q t}\right)
-n_q \left|e^{i\omega_q \tau_{12}}\left(2-e^{i\omega_q t}\right)-1\right|^2 \Bigg]  \Bigg\}\ , \notag
\end{eqnarray}
with $t,\ \tau_{12}>0$. Here, a frame that rotates with the exciton's transition energy was chosen. The exciton-phonon coupling matrix element for the usual deformation potential coupling reads
\begin{equation}
g_{\bf q} = \frac{q}{\sqrt{2\rho \hbar V \omega_q}}\left[
D_{\rm e}e^{ -q^2 a^2_{\rm e}/4 }
 - D_{\rm h}e ^{- q^2 a^2_{\rm h}/4}
\right]\ ,
\end{equation}
where the mass density $\rho$ and the deformation potentials for electrons and holes $D_{\rm e}$ and $D_{\rm h}$, respectively, are material parameters. It has been found that such a spherical parametrization with the localization lengths $a_{\rm e}$ and $a_{\rm h}$ taken as suitable fit parameters also well reproduces the phonon coupling of QDs with an anisotropic shape~\cite{luker2017phon}.
\\
Converting the summation over the phonon wave vectors ${\bf q}$ into a spectral integration by the definition of the phonon spectral density $J(\omega_{\rm ph}) = \int |g_q|^2 \delta(\omega_{\rm ph} - \omega_q) {\rm d}^3q$
and taking the temperature $T=0$, which leads to a vanishing initial Bose distribution of the phonons $n_q=0$ the FWM polarization becomes
\begin{eqnarray}
p_{\rm FWM}(t,\tau_{12}) &\mkern-15mu \sim&\mkern-15mu  \exp\Bigg\{ \int \frac{J(\omega_{\rm ph})^2}{\omega_{\rm ph}^2}
\Big[e^{i\omega_{\rm ph} t} + e^{-i\omega_{\rm ph} t} - 3 \label{eq:pFWM}\\
&&\qquad\qquad + e^{i\omega_{\rm ph}\tau_{12}} \left(2-e^{i\omega_{\rm ph} t}\right) \Big] {\rm d}\omega_{\rm ph}\Bigg\} \ . \notag
\end{eqnarray}
The Fourier transform of this polarization with respect to the real time after the second pulse $t$ gives the phonon sidebands of the FWM spectrum for a given delay $\tau_{12}$. For the special case of $\tau_{12}=0$ the dynamics of the FWM polarization reads
\begin{eqnarray}
p_{\rm FWM}(t,0) &\mkern-15mu \sim&\mkern-15mu  \exp\Bigg[ \int \frac{J(\omega_{\rm ph})^2}{\omega_{\rm ph}^2}
\Big(e^{i\omega_{\rm ph} t}  - 1 \Big) {\rm d}\omega_{\rm ph}\Bigg] \ , \label{eq:p_t_0}
\end{eqnarray}
which directly shows that only spectral features at positive
energies are expected when the two laser pulses hit the QD at the
same time. From \eqref{eq:pFWM} the phonon contribution is isolated by subtracting 
the constant long time value $p_{\rm FWM}(t\to\infty,\tau_{12})$. 
An exemplary delay resolved simulation of this
background is shown in Fig.~\ref{fig:1}(e), where the delay
increases from $\tau_{12}=0$ at the top to the bottom. We find that
the spectrum for most $\tau_{12}$ consists of two maxima located
energetically below and above the zero phonon line (ZPL), i.e., the
exciton transition. Because for positive delays the FWM signal is a measure for the 
coherence in the system the phonon sidebands can be seen as
phonon assisted coherences of the system. We focus first on the vanishing delay case and
clearly see that the phonon sideband only appears on the high energy
side of the ZPL, as already predicted from \eqref{eq:p_t_0}. Because
we consider a vanishing temperature initially no phonons are present
($n_q=0$) that could be absorbed. Therefore we know that this
sideband stems from phonon emission processes. Because the FWM
experiment for a vanishing delay is a photon absorption process this
phonon sideband appears at the high energy side of the ZPL. It is
known that this phonon emission leads to the generation of a wave packet
leaving the QD at the speed of sound~\cite{wigger2013fluc}. Once
this acoustic pulse has left the region of the QD its phonons cannot
be absorbed during optical excitations of the exciton, meaning that
they cannot give rise to sidebands in the FWM spectrum. Therefore
only the generated polaron leads to the remaining sidebands for
large $\tau_{12}$, at the bottom of Fig.~\ref{fig:1}(e). 
We find that these sidebands now appear symmetrically on both sides of
the ZPL and are much less pronounced than the sidebands at short
delays. This can be interpreted as remaining phonon assisted 
coherence in the polaron state. Following the phonon absorption sideband, energetically
below the exciton, back towards short delays we see that also this
process becomes more pronounced before it vanishes at $\tau_{12}=0$.
From the general dynamical behavior of the phonon sidebands we can
conclude that significant variations appear during the polaron
generation, \emph{i.e.}, during the phonon wave packet emission.
Consequently, while the traveling acoustic pulse is still
overlapping with the QD the entire phonon background is increased
promoting not only reabsorption processes, but also stimulated
phonon emission.\\
In Refs.~\cite{vagov2003impa,krugel2007moni} the aforesaid behavior of the phonon
sidebands was already calculated and additional calculations for
increased temperatures predicted a symmetrization of the sidebands
with respect to the phonon emission and absorption side of the ZPL.
To confirm this prediction we perform the first measurement of the
phonon sideband dynamics during the polaron formation process
resolved by FWM spectroscopy. Figure~\ref{fig:2} shows the
measured FWM spectra dynamics in the left column for the
temperatures 5~K, 20~K, and 30~K, in (a), (b), and (c),
respectively. Note that the experiment at 30~K was performed on a
different dot in the same sample, which might lead to some variations of
the geometry of the QDs, although, these are estimated to be
weak~\cite{borri2002rabi}, especially due to the sample
annealing, which is known to decrease the fluctuations of size and Indium
distribution within the ensemble of InGaAs QDs~\cite{braun2016impa}. The
strongest signal in these QD systems naturally defines the ZPL and
therefore appears at the exciton transition with $E-E_{\rm ZPL}=0$.
The detected phonon sidebands spread over a few meV and disappear
within around 2~ps. The laser pulses used for the excitation of the
system have a duration of approximately 150~fs. In the detected
spectral dynamics we directly see that the polaron formation
dynamics happens mainly within approximately 1~picosecond. In this
context simulations in the limit of delta-pulses would not be
justified because the laser pulses are not sufficiently more rapid
with respect to the phonon dynamics. Therefore, to account for
finite pulse durations we simulate the optically driven coupled
exciton-phonon dynamics in a density matrix approach and use a
correlation expansion approach to determine the FWM signal. This
approach has been successfully applied before to model the phonon influence on FWM
signals~\cite{wigger2017expl,wigger2018rabi}.
\\
The results of the simulations with a pulse duration of 150~fs are plotted in the middle column of Fig.~\ref{fig:2} showing a good agreement with the measured data for the three respective temperatures. Both the timescale of the polaron formation, identified by the disappearance of the dominant part of the phonon sidebands (yellow areas), and the energetic spread of the sidebands agree very well. This agreement has been achieved with the localization lengths $a_{\rm e}=4$~nm and $a_{\rm h}=0.87 a_{\rm e}=3.5$~nm and the deformation potentials $D_{\rm e} = 10.5$~eV and $D_{\rm h}=5.25$~eV. Note that because of the rather similar values of $a_{\rm e}$ and $a_{\rm h}$ the phonon coupling is also well reproduced by a standard phonon spectral density with coupling strength $A=0.081$~ps$^2$ and a cut-off frequency $\omega_{\rm c}=1.8$~ps$^{-1}$~\cite{reiter2014role}. We want to remark the values of $a_{\rm e/h}$ in general do not reflect the real dimensions of the QD but act as a spherical representation leading to the same phonon coupling. As a consequence, also $D_{\rm e}$ and $D_{\rm h}$ deviate from the typically used ones by a factor of 1.5~\cite{krummheuer2002theo}. Nevertheless similar to the discussion in Ref.~\cite{jakubczyk2016impa} other choices would not give reasonable agreement, both in the spectral and the time domain. Smaller values for $D_{\rm e/h}$ could in principle be compensated by smaller $a_{\rm e/h}$, as these would increase the coupling strength. But in turn the spectral spread of the phonon sidebands would increase and the polaron formation time would shorten. Thus the agreement between simulation and experiment would be less good.\\
Despite the qualitative agreement, it is not straightforward to compare false-color plots quantitatively. To retrieve a quantity that can directly be compared we plot the spectrally integrated FWM amplitude as a function of the delay in the right column. While the rise of the signal around $\tau_{12}=0$ is mainly given by the laser pulse duration, the following decay directly reflects the loss of coherence due to the phonon wave packet emission during the polaron formation. These dynamics are therefore called phonon induced dephasing of the exciton. For all considered temperatures the curves from experiment (orange) and from theory (blue) agree almost perfectly. Therefore we can conclude that the geometry of the investigated QDs are almost equal.\\
As mentioned before, we expect the strongest asymmetry between the
phonon emission and absorption sidebands for small temperatures. For
larger temperatures more and more phonon modes are already occupied
before the first laser pulse and the ratio between absorption $\sim
n_q$ and stimulated emission $\sim (n_q+1)$ becomes more balanced.
This is what we also find here. When focusing on the $T=5$~K example
in Fig.~\ref{fig:2}(a) we find a stronger sideband at energies
above the ZPL for delays around $\tau_{12}=0.5$~ps. This is in
agreement with the $T=0$ example in Fig.~\ref{fig:1}(e). 
To confirm this asymmetry in Fig.~\ref{fig:3}(a) we plot the difference between the
phonon emission and the absorption side of the FWM spectrum for $T=5$~K.
The left side of the plot shows the experiment and the right side the simulation.
We find significant positive values in both cases until $\tau_{12}=2$~ps predominantly around
$\tau_{12}=0.5$~ps. This confirms that the high energy side of the ZPL has
stronger signals then the respective ones below the ZPL.
When increasing the temperature via $T=20$~K in Fig.~\ref{fig:2}(b)
to $T=30$~K in (c) the sidebands get more and more symmetric because
the contributions stemming from occupied phonon modes $\sim n_q$
overweigh the spontaneous processes $\sim 1$. To also confirm this
symmetrization Fig.~\ref{fig:3}(b) shows the difference spectra for $T=30$~K.
For the entire delay range the measured signal difference shows equal 
positive and negative values, which proves that the phonon emission and
absorption are balanced. This is in good agreement with the simulated 
signal difference.
\\
In conclusion, we have provided the first demonstration of the polaron formation in a single semiconductor QD detected in FWM spectroscopy. In agreement with simulations this process takes a few picoseconds and manifests in characteristic dynamics of phonon sidebands. Also the temperature of the system influences details of the detected sidebands because the occupation of the contributing phonon modes strongly influences the strength of phonon absorption and emission probabilities.\\
Our work shows that experimental and theoretical insight into coherent dynamics of phonon assisted processes in nanostructures is now at hand.  It paves the way toward understanding of more involved phononic systems, for example, exhibiting confined phonon modes~\cite{krummheuer2005coup,lindwall2007zero} or phonon state preparation~\cite{auffeves2014opti}.
\noindent\textbf{Funding.} TK, DW \& PM acknowledge support form the Polish National Agency for Academic Exchange under the International Academic Partnerships program, the W\"urzburg group by the State of Bavaria and C.S by the DFG (project Schn1376-5.1).\\
\noindent\textbf{Disclosures.} The authors declare no conflicts of interest.

\bibliography{PSB}

\bibliographyfullrefs{PSB}

\end{document}